\documentclass[review,authoryear, 12pt]{elsarticle}
\usepackage{epsfig,graphicx,setspace,tabularx,xr}
\usepackage{amsmath,amsthm,amssymb,enumerate}
\usepackage{mathrsfs,mathabx}
\usepackage{bm,verbatim,subfigure}
\usepackage{graphicx}
\usepackage{setspace,color}
\usepackage{multirow}
\usepackage{multicol}
\usepackage{rotating}
\usepackage{threeparttable}
\usepackage{xr,enumitem,CJK}
\usepackage[titletoc]{appendix}
\usepackage{titlesec}
\usepackage{accents}
\usepackage[authoryear]{natbib}

\def\0{\bf{0}}
\def\1{\bf{1}}

\def\bb{{\boldsymbol{\beta}}}
\def\bgamma{{\boldsymbol{\gamma}}}

\def\bpi{{\boldsymbol{\pi}}}

\def\btau{{\boldsymbol{\tau}}}

\def\bepsilon{{\boldsymbol{\epsilon}}}
\def\btheta{{\boldsymbol{\theta}}}

\def\bPsi{{\boldsymbol{\Psi}}}
\def\bTheta{{\boldsymbol{\Theta}}}

\def\bY{\textbf{Y}}
\def\bW{\textbf{W}}

\def\v{\textbf{v}}
\def\br{\textbf{r}}

\def\bse{\begin{eqnarray*}}
\def\ese{\end{eqnarray*}}
\def\be{\begin{eqnarray}}
\def\ee{\end{eqnarray}}
\def\bsq{\begin{equation*}}
\def\esq{\end{equation*}}
\def\bq{\begin{equation}}
\def\eq{\end{equation}}

\def\var{\hbox{var}}

\def\wh{\widehat}
\def\wt{\widetilde}
\def\wc{\widecheck}

\def\n{\nonumber}

\def\diag{{\rm diag}}

\def\sumi{\sum_{i=1}^n}

\def\sumj{\sum_{j=1}^m}
\def\sumk{\sum_{k=1}^m}
\def\sums{\sum_{s=1}^m}
\def\sumt{\sum_{t=1}^m}
\def\sumq{\sum_{q=1}^m}
\def\trans{^{\rm T}}

\def\bTheta{\boldsymbol\Theta}

\def\0{{\bf 0}}

\def\a{{\bf a}}

\def\V{{\bf V}}
\def\g{{\bf g}}
\def\r{{\bf r}}

\def\u{{\bf u}}
\def\v{{\bf v}}

\def\M{{\bf M}}

\def\u{{\bf u}}

\def\X{{\bf X}}
\def\x{{\bf x}}

\def\R{{\bf R}}

\def\Z{{\bf Z}}

\def\bPhi{{\boldsymbol \Phi}}

\def\bq{\begin{equation}}
\def\eq{\end{equation}}
\def\pr{\hbox{pr}}

\def\log{\hbox{log}}

\def\squarebox#1{\hbox to #1{\hfill\vbox to #1{\vfill}}}
\def\btheta{{\boldsymbol \theta}}

\def\bpi{{\boldsymbol \pi}}

\def\var{\hbox{var}}

\def\pr{\hbox{pr}}
\def\log{\hbox{log}}

\def\boxit#1{\vbox{\hrule\hbox{\vrule\kern6pt\vbox{\kern6pt#1\kern6pt}\kern6pt\vrule}\hrule}}

\def\tbPhi{\wt{\boldsymbol{\Phi}}}
\def\tPsi{\wt{\boldsymbol{\Psi}}}
\def\tv{\wt{\textbf{v}}}
\def\tM{\wt{\textbf{M}}}
\def\tu{\wt{\textbf{u}}}
\def\tv{\wt{\textbf{v}}}
\def\tV{\wt{\textbf{V}}}
\def\tg{\wt{\textbf{g}}}

\def\tbtheta{{\wt{\boldsymbol{\theta}}}}

\def\cbtheta{{\wc{\boldsymbol{\theta}}}}
\def\cg{\wc{\textbf{g}}}
\def\cbPhi{\wc{\boldsymbol{\Phi}}}

\def\cv{\wc{\bf v}}
\def\cM{\wc{\bf M}}
\def\cu{\wc{\bf u}}
\def\cv{\wc{\bf v}}
\def\cV{\wc{\bf V}}

\newcommand{\tabincell}[2]{\begin{tabular}{@{}#1@{}}#2\end{tabular}}

\newtheorem{Th}{\underline{\bf Theorem}}
\newtheorem{Alg}{\underline{\bf Algorithm}}

\newtheorem{Rem}{\underline{\bf Remark}}

\allowdisplaybreaks

\journal{}

\begin{document}
\begin{frontmatter}
\title{Semiparametric Mixture Regression with Unspecified Error Distributions}

\author[address1]{Yanyuan Ma\footnote{Ma's research is supported by
    National Sciences Foundation and National Institute of Neurological
    Disorders and Strokes.}}
\author[address2]{Shaoli Wang}
\author[address3]{Lin Xu\footnote{Xu's research is supported by Zhejiang Federation of Humanities and Social Sciences Circles grant 16NDJC154YB.}}
\author[address4]{Weixin Yao\footnote{corresponding author. weixin.yao@ucr.edu. Yao's research is supported by NSF grant DMS-1461677 and and Department of Energy with the award DE-EE0007328.}}

\address[address1]{Department of Statistics, The Pennsylvania State University}
\address[address2]{School of Statistics and Management, Shanghai University of Finance and Economics}
\address[address3]{School of Data Sciences, Zhejiang University of Finance and Economics}
\address[address4]{Department of Statistics, University of California, Riverside}

\begin{abstract}
In fitting a mixture of linear regression models, normal
assumption is traditionally used to model the error and then
regression parameters are estimated by the maximum likelihood
estimators (MLE). This procedure
is not valid if the normal assumption is violated. To relax the
normal assumption on the error distribution hence reduce the modeling bias,  we
propose semiparametric mixture of linear regression models
with unspecified error distributions.
We establish a more general identifiability result under weaker
conditions than existing results, construct a class of new estimators,
and establish their asymptotic properties.
These asymptotic results also apply to many existing semiparametric
mixture regression estimators whose asymptotic properties have
remained unknown due to the inherent difficulties in obtaining them.
Using simulation studies, we demonstrate
the superiority of the proposed estimators over the MLE when the normal
error assumption is violated and the
comparability when the error is normal. Analysis of a newly collected
Equine Infectious Anemia
Virus  data in 2017 is employed to illustrate the usefulness of the
new estimator.
\end{abstract}

\end{frontmatter}
\doublespacing

\section{Introduction}

Mixtures of regressions provide a flexible tool to investigate the
relationship between variables coming from several
unknown latent components. It is widely used in many fields, such
as engineering, genetics, biology, econometrics and marketing. 
A typical mixture of regressions model is as
follows. Let $Z$ be a latent class indicator with
$\pr(Z=j\mid \X)=\pr(Z=j)=\pi_j$
 for $j=1,2,\cdots,m$, where
$\X$ is a $(p+1)$-dimensional vector with the first component the
constant 1 and the rest random predictors. Given $Z=j$,
the response $Y$ depends on $\X$ through  a linear regression model
\begin{equation}
Y=\X\trans\bb_j+\epsilon_j, \label{mixlin}
\end{equation}
where $\bb_j=(\beta_{0j},\beta_{1j},\ldots,\beta_{pj})\trans$, and
$\epsilon_j\sim N(0,\sigma_j^2)$ is independent of $\X$.
Thus the conditional density of $Y$ given $\X=\x$
can be written as
\begin{eqnarray}
f_{Y\mid\X}(y,\x)=\sum_{j=1}^{m}\pi_{j}\phi(y;\x\trans\bb_{j},\sigma_{j}^{2}),
\label{mixreg}
\end{eqnarray}
where $\phi(\cdot;\mu,\sigma^2)$ is the normal density with mean $\mu$ and variance $\sigma^2$.
The unknown parameters in model (\ref{mixreg}) can be estimated
by the maximum likelihood estimator (MLE) using the EM algorithm
\citep{dempster77}. See, for example, \cite{Wedel00},
\cite{Skrondal04}, \cite{Jacobs91} and \cite{Jiang91} for some applications of model
(\ref{mixreg}).

A major drawback of model (\ref{mixreg}) is the  normal
assumption of the error density, which does not always hold in practice.
Unfortunately, unlike the equivalence between the
MLE and the least squares estimator
(LSE) in linear regression, the normal assumption of $\epsilon$ in
(\ref{mixlin}) is indispensible for the consistency of  MLE. Furthermore,
the normal assumption is also critical for the
computation of MLE because it is needed when calculating the classification
probabilities in the E step of the EM algorithm.

In order to reduce the modeling bias, we
relax the normal assumption of the component error
distributions and propose a class of flexible
semiparametric mixture of linear regression models by replacing the
normal error densities in (\ref{mixreg}) with unspecified
component error densities. Specifically, we propose a
semiparametric mixture of linear regressions model of the form
\be
f_{Y\mid\X}(y,\x,
\btheta,g)
=\sum_{j=1}^m\pi_j\tau_jg\{(y-\x\trans\bb_j)\tau_j\}, \label{newmodel}
\ee
where $\btheta=(\pi_1,\ldots,\pi_{m-1},\bb_1\trans,\ldots,\bb_m\trans,
\tau_1, \dots, \tau_m)\trans$ and $g$ is an unspecified density
function with mean zero and variance one. Note that
$\pi_m=1-\sum_{j=1}^{m-1}\pi_j$ and we can view
$\tau_j$ as the scale parameter or precision parameter playing the
role of $\sigma_j^{-1}$ in (\ref{mixreg}).  For a
special case of (\ref{newmodel}) where $\tau_1=\tau_2=\cdots=\tau_m$
and $g$ is a symmetric function, some existing work on identifiability
exists. For example,
\citet{Bordes06} and
\citet{Hunter07} established the model identifiability
when $m\le3$ and $\X=1$, i.e. when the regression model
degenerates to a mixture of density functions,
while \cite{Hunter12} allowed any $m$ and included
covariates in $\X$. In this work,
we establish the identifiability result for model
(\ref{newmodel}) in a more general setting than the existing literatures, where the
identifiability is shown for the arbitrary component densities $g_j$ with mean 0 without the identical constraint
on the $\tau_j$'s. We also propose a semiparametric EM
algorithm to estimate the regression parameters $\btheta$ and the
unspecified function $g$. We further prove
the consistency and the asymptotic properties of the new semiparametric
estimator. Our asymptotic results directly apply to many existing
semiparametric mixture regression estimators whose asymptotic
properties have not been established in the literature. Using a Monte Carlo simulation
study, we demonstrate
that our methods perform better than the traditional MLE when the
errors have distributions other than normal and provide
comparable results when the
errors are normal.
An empirical analysis of a newly collected Equine Infectious Anemia Virus (EIAV)
data set in 2017 is carried out to  illustrate the usefulness of the proposed
methodology.

The rest of the paper is organized as follows. Section
\ref{sec:newmethod} introduces the new mixture of regressions model with
unspecified error distributions, proposes the new semiparametric
regression estimator, and establishes the asymptotic properties of
the proposed estimator. In Section \ref{sec:Simulation}, we use a
simulation study to demonstrate  the superior performance of the new
method. We illustrate the effectiveness of
the new method on an
EIAV data set in Section \ref{sec:Realdata}. Some discussions are given
in Section \ref{sec:summary}.

\section{Mixture of regressions with nonparametric error densities}
\label{sec:newmethod}
\subsection{Identifiability results}

Before proposing estimation procedures, we first
  investigate the identifiability of  the model in
(\ref{newmodel}).
Let
$\X=(1,\X_s\trans)\trans,\bb_j=(\beta_{0j},\bb_{sj}\trans)\trans$ with
$\bb_{sj}=(\beta_{1j},\ldots,\beta_{pj})\trans$.
\begin{Th} \label{th:identifiable}
(Identifiability) Assume that
$\pi_j>0$, $j=1,\cdots,m$, and $\bb_{sj}$, $j=1,\cdots,m$, are
distinct vectors in ${\mathbf R}^{p}$. Assume further that the support
of $\X_s$ contains an open set in ${\mathbf R}^p$. Then the semiparametric mixture regression model
(\ref{newmodel}) is identifiable up to a permutation of the $m$ components.
\end{Th}

\begin{Rem}
A more general identifiability result is proved in the supplementary
document. More specifically, under the assumptions in Theorem
\ref{th:identifiable}, the model
\begin{equation}
f_{Y\mid\X}(y,\x,\btheta,\g)=\sum_{j=1}^m\pi_jg_j(y-\x\trans\bb_j),
\label{newmodel1}
\end{equation}
is identifiable, where $g_j$ has mean 0 and $\g=(g_1,\ldots,g_m)$. Note that model
(\ref{newmodel}) is a special case of (\ref{newmodel1})
when $g_j(\cdot)$'s belong to the same distribution family with different
precision parameters.
Our identifiability result benefits from
the information carried in the random covariates $\X_s$.
This allows us to establish the identifiability
result for general number of components $m$ and arbitrary $g_j(\cdot)$'s.
\end{Rem}

\subsection{Estimation algorithms}
\label{sec:alg}

Suppose that $\{(\X_1,Y_1),\ldots,(\X_n,Y_n)\}$ are random
observations from (\ref{newmodel}). In this section, we
propose a Kernel DEnsity based EM type algorithm (KDEEM) to estimate
the parameter $\btheta$ and the nonparametric density function
$g(\cdot)$ in  (\ref{newmodel}):
\begin{Alg} \label{alggscalemle}{\rm
Starting from an initial parameter $\btheta^{(0)}$ and initial density
function $g^{(0)}(\cdot)$, at the $(k+1)$th step,
\begin{description}
\item[E step:]  Calculate the classification probabilities,
\[p_{ij}^{(k+1)}=\pr(Z_i=j\mid\x_i,y_i)=\frac{\pi_j^{(k)}g^{(k)}(r_{ij}^{(k)})\tau_j^{(k)}}{\sum_{j=1}^m\pi_j^{(k)}g^{(k)}(r_{ij}^{(k)})\tau_j^{(k)}},\:i=1,\ldots,n,\:
j=1,\ldots,m,\] where $\epsilon_{ij}^{(k)}=y_i-\x_i\trans\bb_j^{(k)}$ and $r_{ij}^{(k)}=\epsilon_{ij}^{(k)}\tau_j^{(k)}$.
\item[M step:] Update $\btheta$ and $g(\cdot)$, via
\begin{enumerate}
\item
$\pi_j^{(k+1)}=n^{-1}\sumi p_{ij}^{(k+1)}$,
\item
$
({\bb_j^{(k+1)}}\trans,
{\tau_j^{(k+1)}})\trans=\arg\max_{\bb_j,\tau_j}
\sumi p_{ij}^{(k+1)}\log
[g^{(k)}\{(Y_i-\x_i\trans\bb_j)\tau_j\}\tau_j]$, 
for $ j=1, \dots,m$.
\item
$
g^{(k+1)}(t)=n^{-1}\sum_{i=1}^n\sum_{j=1}^mp_{ij}^{(k+1)}K_h(r_{ij}^{(k+1)}-t),
$
where  $j=1,\ldots,m$,
$K_h(t)=h^{-1}K(t/h),$ and $K(t)$ is a kernel function, such as
the Epanechnikov kernel.
\end{enumerate}
\end{description}}
\end{Alg}
For the conventional MLE, the normal density for $g(\cdot)$ is
used to calculate the classification probabilities in the E step. In
the KDEEM, the error density used in the E
step is estimated by a weighted kernel density estimator in stage 3 of
the M step, with classification probabilities as weights, to
avoid the modelling bias of component error
densities. \cite{Bordes07} and \cite{Benaglia09} have used similar idea
of combining
kernel density and EM algorithm for the mixture of location shifted
densities when there are no predictors involved.  Note
  that the above EM type algorithm cannot guarantee to increase the
  likelihood at
each iteration due to the kernel density estimation in the M step. One
could use the maximum smoothed loglikelihood method proposed by
\cite{levine2011maximum} to produce a modified algorithm that does
increase smoothed version of the loglikelihood at each iteration but
provides similar performance to the KDEEM.

\cite{Hunter12} considered a special case of Algorithm
\ref{alggscalemle} by assuming homogeneous scales, i.e.
$\tau_1=\tau_2=\cdots=\tau_m$, denoted by KDEEM.H. For completeness of
the presentation, we also present out the EM algorithm for this special
case.

\begin{Alg} \label{alggsamemle}{\rm
Starting from an initial parameter $\btheta^{(0)}$ and initial density
function $g^{(0)}(\cdot)$, at the $(k+1)$th step,
\begin{description}
\item[E step:]  Calculate the classification probabilities,
\bse
p_{ij}^{(k+1)}=P(\Z_i=j\mid \x_i,
y_i)=\frac{\pi_j^{(k)}g^{(k)}(\epsilon_{ij}^{(k)})}{\sumj\pi_j^{(k)}g^{(k)}(\epsilon_{ij}^{(k)})}\:i=1,\ldots,n,\:
j=1,\ldots,m,
\ese where $\epsilon_{ij}^{(k)}=y_i-\x_i\trans\bb_j^{(k)}$.
\item[M step:] Update $\btheta$ and $g$, via
\begin{enumerate}
\item
$\pi_j^{(k+1)}=n^{-1}\sumi p_{ij}^{(k+1)}$,
\item
 $\bb_j^{(k+1)}=\arg\max_{\bb_j}\sumi p_{ij}^{(k+1)}\log
[g^{(k)}\{(Y_i-\x_i\trans\bb_j)\}]$, 
for $ j=1, \dots,m$.
\item
\begin{equation}\label{gsame}
  g^{(k+1)}(t)=n^{-1}\sum_{i=1}^n\sum_{j=1}^mp_{ij}^{(k+1)}K_h(\epsilon_{ij}^{(k+1)}-t),
\end{equation}
where  $j=1,\ldots,m$.
\end{enumerate}
\end{description}}
\end{Alg}

To simplify the computation, \cite{Hunter12} also recommended to use the
least squares criterion to update $\bb$ in the M step of Algorithm
\ref{alggsamemle}, i.e.,
\bse
\bb_j^{(k+1)}=(\X\trans\bW_j^{(k+1)}\X)^{-1}\X\trans\bW_j^{(k+1)}\bY,
\ese
where  $\X=(\x_1,\ldots,\x_n)\trans,\bY=(y_1,\ldots,y_n)\trans,
\bW_j^{(k+1)}=\diag\{p_{1j}^{(k+1)},\ldots,p_{nj}^{(k+1)}\}$. Let
$\cbtheta$ and $\cg(\cdot)$ be the resulting estimators, denoted by
KDEEM.LSE.  Note that
$\cbtheta$ is different from the classic MLE in that
the classification probabilities are calculated based on the weighted
kernel density estimator (\ref{gsame}) instead of the normal density
to avoid the misspecification of the component error densities.

\subsection{Asymptotic properties}
We now establish the asymptotic properties of the
  estimators presented in Section
\ref{sec:alg}. Let $\wh{\btheta}$ and $\wh{g}(\cdot)$ be the resulting
estimators of Algorithm
\ref{alggscalemle} and $\btheta_0$ and $g_0(\cdot)$ be the corresponding true
values. Next,
we provide
the asymptotic properties of $\wh{\btheta}$ and $\wh{g}(\cdot)$. We make the
following mild
Assumptions.
\begin{enumerate}[label=A\arabic*]

\item\label{assum:pi} The probabilities $\pi_j\in(0,1)$ for $j=1, \dots,
  m$, $\sumj\pi_j=1$.

\item\label{assum:tau} The precision values satisfy $0<\tau_j<\infty$ for
  $j=1, \dots, m$.

\item\label{assum:beta} The true parameter value is in the interior of
  an open set $\bTheta\subset{\cal R}^d$ where $d=\dim(\btheta)$.

\item\label{assum:g} The pdf $g(\cdot)$ has a compact support and is
bounded away from zero on its support. In addition, $g(\cdot)$ is
continuous and has continuously bounded second derivative with mean 0 and variance 1.

\item\label{assum:kernel} The kernel function $K(\cdot)$ is symmetric,
  bounded, and twice differentiable with bounded second derivative, compact support and finite second moment.

\item\label{assum:h} The bandwidth $h$ satisfies $nh^2\to\infty$ and
  $nh^4\to0$ when $n\to\infty$.

\item\label{assum:h} In the neighborhood of the true
    parameter values ($\btheta_0$, $g_0$), there is a
    unique value ($\wh\btheta$, $\wh g$) where the EM algorithm
    converges to.

\end{enumerate}

To state the theoretical results in Theorem \ref{th:diffg}, we first
define some notations, while collect the proof of Theorem
\ref{th:diffg} in Section
 S.2 of the Supplementary document.

 For any vector $\a=(a_1,\ldots,a_p)$, let $g(\a)$ be the element-wise evaluation of $g(\cdot)$ at $\a$. Let $\br_i=(r_{i1},\ldots,r_{im}),$ where $r_{ij}=(y_i-\x_i\trans\bb_j)\tau_j$.
Define
\begin{align*}
\bPhi\{\x_i, y_i, g(\br_{i}),\btheta\}
&=\left[
\begin{array}{c}
\bPhi_1\{\x_i, y_i,g(\br_{i}),\bpi,\btau,\bb_1\}\\
\vdots\\
 \bPhi_m\{\x_i, y_i,g(\br_{i}),\bpi,\btau,\bb_m\}
\end{array}\right],
\end{align*}
where $\bpi=(\pi_1,\ldots,\pi_{m-1})\trans,\btau=(\tau_1,\ldots,\tau_m)\trans$, and
\begin{align*}
&\bPhi_j\{\x_i,y_i,g(\br_{i}),\bpi,\btau,\bb_j\}=\left[
\frac{g(r_{ij})\tau_j}{\sumj\pi_jg(r_{ij})\tau_j}-1,
\frac{\pi_j\{g(r_{ij})+g'(r_{ij})r_{ij}\}}{\sumj\pi_jg(r_{ij})\tau_j},
\frac{\pi_jg'(r_{ij}) \tau_j\x_i\trans}{\sumj\pi_j
    g(r_{ij})\tau_j}
\right]\trans.
\end{align*}
Also, let
\bse
\Psi\{t, g(t),g(\br_{i}),\btheta\}
=\frac{\sumj \pi_j
 g(r_{ij})\tau_j
K_h(r_{ij}-t)}{\sumj\pi_j
g(r_{ij})\tau_j}- g(t).
\ese
Let
$g(\cdot, \btheta)$ satisfy
$E[\Psi\{t,g(t), g(\br_{i}),\btheta\}]=0$ for all $\btheta$ and $t$. Define
\bse
\r_{21}(t)&=&E\left[
\frac{\partial\Psi\{t,g(t), g(\br_{i}),\btheta\}}
{\partial\btheta}\right]\bigg|_{g(\cdot)=g(\cdot, \btheta)},\\
r_{22}(t)&=&E\left[\frac{\partial\Psi
\{t,g(t,\btheta), g(\br_{i},\btheta),\btheta\}}{\partial
g(t,\btheta)}\right],\\
\R_i(t)&=&\left[\begin{array}{c}
{\partial\Psi
\{t,g(t,\btheta), g(\br_{i},\btheta),\btheta\}}/{\partial
                  g(r_{i1},\btheta)}\\
\vdots\\
{\partial\Psi
\{t,g(t,\btheta), g(\br_{i},\btheta),
\btheta\}}/{\partial
                  g(r_{im},\btheta)}
\end{array}\right].
\ese
Further let $\r_{21}(\x_i, y_i)=\{\r_{21}(r_{i1}),
\dots,\r_{21}(r_{im})\}$,
$\r_{22}(\x_i, y_i)=\diag\{r_{22}(r_{i1}),
\dots,r_{22}(r_{im})\}$,
$\r_{21}=\{\r_{21}(\x_1, y_1), \dots, \r_{21}(\x_n,
y_n)\}$,
$\r_{22}=\diag\{\r_{22}(\x_1, y_1), \dots,
\r_{22}(\x_n,y_n)\}$. Also let
$\R_l(\x_i,y_i)=\{\R_l(r_{i1}),\dots, \R_l(r_{im})\}$,
 $\R_i=\{\R_1\trans(\x_i,y_i), \dots,
\R_n\trans(\x_i,y_i)\}\trans$,
$\R=(\R_1, \dots, \R_n)$. Let
\[\r_3(\x_i, y_i)=\left[\frac{\partial \bPhi\{\x_i, y_i, g(\br_{i},\btheta),\btheta\}}{\partial
g(r_{i1},\btheta)},
\dots,
\frac{\partial \bPhi\{\x_i, y_i, g(\br_{i},\btheta),\btheta\}}{\partial
g(r_{im},\btheta)}\right],\]
and write $\r_3=\{\r_3(\x_1, y_1), \dots, \r_3(\x_n, y_n)\}$.
Now we define
 \bse
\M=E\left\{n^{-1}\r_3
(\r_{22}+n^{-1}\R)^{-1}\r_{21}\trans\right\}
-E\frac{\partial
\bPhi\{\X_i, Y_i, g(Y_i-\X_i\trans\bb_1), \dots,
g(Y_i-\X_i\trans\bb_m),\btheta\}
}{\partial\btheta\trans}.
\ese
Define also
\bse
\u(\x_i,y_i)&=&\sumk
\sumj
E\left[
\left\{\frac{\sigma_j\pi_k\tau_kg(r_{ik},\btheta)}
{\sums\pi_s \tau_sg(r_{is},\btheta)}\right\}
\left\{\sums
    \pi_s\tau_sg(\gamma_{ijks})\right\}
\right.\n\\
&&\left.\times
\frac{\partial\bPhi
\{\X, \X\trans\bb_j+\sigma_jr_{ik}, g(\bgamma_{ijk}),\btheta\}
}{\partial g(r_{ik},\btheta)}
\right]-\sumk
E\left[\frac{\partial\bPhi
\{\X, Y, g(\br_{1},\btheta),\btheta\}
}{\partial g(r_{1k},\btheta)}g(r_{1k},\btheta)
\right],
\ese
and
\bse
\v(\x_i,y_i)&=&
\sumj\sumk
E\left[\frac{\partial\bPhi
\{\X, \sigma_jr_{ik}+\X\trans\bb_j, g(\bgamma_{ijk}), \btheta\}
}{\partial g(r_{ik},\btheta)}\pi_k\tau_k \sigma_j\frac{
f_{Y\mid\X}(\sigma_jr_{ik}+\X\trans\bb_j,\X)
 g(r_{ik})
}{\{\sums\pi_s\tau_s g(r_{is})\}}\right]\n\\
&&\hspace{-1.5cm}-
\sumj\sumk\sums\sumt
E\left[\frac{\partial\bPhi
\{\X_1, \eta_{ijkst},g(\tau_1(\eta_{ijkst}-\X_1\trans\bb_1)),\dots,g(\tau_m(\eta_{ijkst}-\X_1\trans\bb_m)),\btheta\}}
{\partial g(\tau_s(\sigma_kr_{it}+\X_l\trans\bb_k-\X_l\trans\bb_s))}\right.\n\\
&&\hspace{-1.3cm}\left.\times\frac{g(\tau_s(\sigma_kr_{it}+\X_2\trans\bb_k-\X_2\trans\bb_s))\sumq\pi_q\tau_qg(\tau_q(\eta_{ijkst}-\X_1\trans\bb_q)
)}{\{\sumq\pi_q\tau_q
  g(\tau_q(\sigma_kr_{it}+\X_2\trans\bb_k-\X_2\trans\bb_q))\}}\right]\frac{\tau_s\sigma_j\pi_k\pi_s\pi_t\tau_tg(r_{it})}{\{\sumq\pi_q\tau_q
  g(r_{iq})\}}\n,
\ese
where $\sigma_j=\tau_j^{-1}, \gamma_{ijkl}=\tau_l(\X\trans\bb_j+\sigma_jr_{ik}-\X\trans\bb_l), \bgamma_{ijk}=(\gamma_{ijk1},\ldots,\gamma_{ijkm}),$ and $\eta_{ijkst}=\sigma_j\tau_s(\sigma_kr_{it}+\X_2\trans\bb_k-\X_2\trans\bb_s)+\X_1\trans\bb_j$.

\begin{Th}\label{th:diffg}
Under the Assumptions \ref{assum:pi}-\ref{assum:h},
the regression
parameter estimator $\wh{\btheta}$ obtained from Algorithm
\ref{alggscalemle} is consistent and satisfies
\bse
\sqrt{n}(\wh\btheta-\btheta_0)\to N(0,\V)
\ese
in distribution when $n\to\infty$,
where
\[\V=\M^{-1}\var
[\bPhi\{\x_i, y_i, g(r_{i1},\btheta), \dots,
g(r_{im},\btheta),\btheta\}+\u(\x_i,y_i)+\v(\x_i,y_i)]{\M^{-1}}\trans.\]
In addition,
$\wh g(t)-g_0(t)=O_p\{h^2+(nh)^{-1/2}\},$ for any $t$.
\end{Th}
Theorem \ref{th:diffg} establishes the theoretical properties of the
estimator $\wh{\btheta}$ in Algorithm
\ref{alggscalemle}. It also shows that
the nonparametric density estimator $\wh g(t)$ has
 the same convergence properties as the classical nonparametric
 density estimator. The proof of Theorem \ref{th:diffg} is lengthy and
 quite involved.

 Let $\wt\btheta$ and $\wt g$ be the resulting estimators
  of Algorithm \ref{alggsamemle} under the assumption of
homogeneous component scales considered in \cite{Hunter12}. In Theorem
\ref{th:samegmle} below, we show the consistency of
$\wt{\btheta}$ and $\wt{g}$
and provide their convergence rate properties, which have remained
unsolved and are considered to be difficult
to obtain in the literature. We benefit from the proof of Theorem
\ref{th:diffg}, which sheds light on the problem and provides the
basic approach to the proof.
We first define some notations, while collect the detailed proof of Theorem
\ref{th:samegmle} in the Supplement.

 Define
\bse
\tbPhi\{\x_i, y_i, g(\bepsilon_{i}),\btheta\}&=&\left[\begin{array}{c}
\tbPhi_1\{\x_i, y_i,g(\bepsilon_{i}),\bpi,\bb_1\}\\
\vdots\\
 \tbPhi_m\{\x_i, y_i,g(\bepsilon_{i}),\bpi,\bb_m\}
\end{array}\right],
\ese
where $\epsilon_{ij}=y_i-\x_i\trans\bb_j$ and
\bse
\tbPhi_j\{\x_i,y_i,g(\bepsilon_{i}),\bpi,\bb_j\}=
\left[
\frac{g(\epsilon_{ij})}{\sumj\pi_jg(\epsilon_{ij})}-1,
\frac{\pi_j \x_i\trans g'(\epsilon_{ij})}{\sumk\pi_k
    g(\epsilon_{ik}) }
\right]\trans.
\ese
Also let
\be
\tPsi\{t, g(t),g(\bepsilon_{i}),\bpi\}
=\frac{\sumj \pi_j
 g(\epsilon_{ij})
K_h(\epsilon_{ij}-t)}{\sumj\pi_j
g(\epsilon_{ij})}- g(t). \label{Psi}
\ee
 Let $\tM, \tu(\x_i,y_i),$ and $\tv(\x_i,y_i)$ be defined similarly in Theorem \ref{th:diffg} by replacing $\{\bPhi(\cdot),\bPsi(\cdot)\}$ with $\{\tbPhi(\cdot),\tPsi(\cdot)\}$ and replacing $r_{ij}$ with $\epsilon_{ij},i=1,\ldots,n,j=1,\ldots,m.$ (See Section S.2 of the Supplementary document for more detail.)

\begin{Th}\label{th:samegmle}
Under the Assumptions \ref{assum:pi}-\ref{assum:h},
$\tbtheta$ is
consistent and satisfies
\bse
\sqrt{n}(\tbtheta-\btheta_0)\to N(0,\tV)
\ese
in distribution when $n\to\infty$,
where
\bse
\tV={\tM}^{-1}\var
[\tbPhi\{\x_i, y_i, g(\epsilon_{i1},\btheta), \dots,
g(\epsilon_{im},\btheta),\btheta\}+\tu(\x_i,y_i)+\tv(\x_i,y_i)]
(\tM^{-1})\trans.
\ese
In addition,
$\tg(t)-g_0(t)=O_p\{h^2+(nh)^{-1/2}\},$ for any $t$.
\end{Th}

Next, we establish the consistency and the asymptotic normality of the ${\cbtheta}$, which is the least squares version of Algorithm \ref{alggsamemle} \citep{Hunter12}. To formally state the theoretical results in Theorem
\ref{th:samegols}, we need to
define some notations, while give the proof of Theorem \ref{th:samegols} in the Supplement.

Let
\bse
\cbPhi\{\x_i, y_i, g(\epsilon_{i1}), \dots, g(\epsilon_{im}),\btheta\}
&=&\left[\begin{array}{c}
\cbPhi_1\{\x_i, y_i,g(\epsilon_{i1}), \dots, g(\epsilon_{im}),\bpi,\bb_1\}\\
\vdots\\
 \cbPhi_m\{\x_i, y_i,g(\epsilon_{i1}), \dots, g(\epsilon_{im}),\bpi,\bb_m\}
\end{array}\right],
\ese
where
\bse
\cbPhi_j\{\x_i,y_i,g(\epsilon_{i1}), \dots, g(\epsilon_{im}),\bpi,\bb_j\}=
\left[
\frac{g(\epsilon_{ij})}{\sumj\pi_jg(\epsilon_{ij})}-1,
\frac{\pi_j\epsilon_{ij}g(\epsilon_{ij}) \x_i\trans}{\sumj\pi_j
    g(\epsilon_{ij})}
\right]\trans.
\ese
Let $\cM, \cu(\x_i,y_i)$, and $\cv(\x_i,y_i)$ be defined similarly in Theorem \ref{th:diffg} by replacing $\{\bPhi,\bPsi\}$ with $\{\cbPhi,\tPsi\}$ and replacing $r_{ij}$ with $\epsilon_{ij},i=1,\ldots,n,j=1,\ldots,m.$ (See the Supplement for more detail.)
\begin{Th}\label{th:samegols}
Under the Assumptions \ref{assum:pi}-\ref{assum:h}, ${\cbtheta}$ is
consistent and satisfies
\bse
\sqrt{n}(\cbtheta-\btheta_0)\to N(0,\cV)
\ese
in distribution, where
\bse
\cV=\cM^{-1}\var
[\cbPhi\{\x_i, y_i, g(\epsilon_{i1},\btheta), \dots,
g(\epsilon_{im},\btheta),\btheta\}+\cu(\x_i,y_i)+\cv(\x_i,y_i)](\cM^{-1})\trans.
\ese
In addition,
$\cg(t)-g_0(t)=O_p\{h^2+(nh)^{-1/2}\},$ for any $t$.
\end{Th}
From the proof of Theorem \ref{th:samegols} in Section
 S.4 of the Supplementary document,
it is easy to see that we can also get consistent mixture
regression parameter estimates if we replace the least squares
criterion in the M step by other robust criteria such as Huber's
$\psi$ function \citep{Huber1981} or Tukey's bisquare function. The
consistency of the estimators can be retained mainly because there is no
modeling misspecification when estimating classification probabilities
in the E step.

\section{Simulation study}
\label{sec:Simulation}

We conduct a series of simulation studies to demonstrate the effectiveness of the
proposed estimator KDEEM under different scenarios of error distributions and
compare them with the traditional normal assumption based MLE via the EM
algorithm (MLEEM). For the proposed estimator, we use the
traditional MLE as the initial values and select the bandwidth
of the kernel density estimation of $g(\cdot)$ based on the method proposed by
\cite{Sheather91}. Better estimation results might be obtained if more sophisticated
methods were used to select the bandwidth. See, for example,
\cite{Sheather91} and \cite{Raykar06}.
For illustration purpose, we also include KDEEM.H presented in
Algorithm \ref{alggsamemle} and the corresponding least squares
version KDEEM.LSE proposed by \cite{Hunter12} for comparison.

We generate the independent and identically distributed (i.i.d.) data
$\{(x_{i},y_{i}), i=1,\ldots, n\}$ from the model
\[Y=\left\{
      \begin{array}{ll}
        -3+3X+\epsilon_1, & \hbox{if $Z=1$;} \\
        3-3X+\epsilon_2, & \hbox{if $Z=2$,}
      \end{array}
    \right.\]
where $Z$ is the component indicator of $Y$ with
$\pr(Z=1)=0.5$, and $X\sim U(0,1)$.

We consider the following cases for the error distribution
$\epsilon_1$ and $\epsilon_2$:
\begin{description}
\item {Case I}: $\epsilon_1\sim N(0,1)$,
\item {Case II}: $\epsilon_1\sim U(-3,3)$,
\item {Case III}: $\epsilon_1\sim 0.5N(-1.5,0.5^2)+0.5N(1.5,0.5^2)$,
\item {Case IV}: $\epsilon_1\sim 0.5N(-1,0.5^2)+0.5N(1,1.5^2)$,
\item {Case V}: $\epsilon_1\sim \Lambda(0,1^2)$,
\item {Case VI}: $\epsilon_1\sim Gamma(2,0.5)$,
\item {Case VII}: $\epsilon_1\sim Rayleigh(3)$,
\end{description}
and  $\epsilon_2$ has the same distribution as
 $0.5\epsilon_1$, i.e.  $\epsilon_2\sim 0.5\epsilon_1$. We
use Case I to check the efficiency loss of the new semiparametric
mixture regression estimators compared to MLE when the error
distribution is indeed normal. The distribution in Case III is  bimodal
and the distribution in Case IV is right skewed. Cases
  II, V, VI, and VII are non-normal error densities and are used to check
  the adaptiveness of the new method to various densities.

In Tables \ref{Table4} to \ref{Table12}, we report the mean absolute
bias (MAB) and the root mean squared error (RMSE) of the regression parameter
estimates based on 1,000 replicates for all seven cases and for
$n=250, 500, 1000$. For convenience of reading, all
the values are multiplied by a factor of $10^2$. In addition, for
better comparison, we also report the relative
efficiency (RE) for each estimator when compared to the classical
method MLEEM. For example, RE of KDEEM is calculated as
\[RE=\left\{\frac{RMSE(MLEEM)}{RMSE(KDEEM)}\right\}^2.\]
A larger value of $RE$ indicates
better performance of the proposed method.
Based on the simulation results, in Case I,
when the error distribution is normal, MLE is the most efficient one
as expected.
However, for Cases II to VII, where the error pdf is not normal,  KDEEM
outperforms MLEEM and the improvement is very substantial, especially for the
slope parameters. In addition, for all cases, KDEEM performs better
than KDEEM.H and KDEEM.LSE, which is expected since the data generation
models have the heterogeneous component scales.

\begin{sidewaystable}[htbp]
\renewcommand{\arraystretch}{1.5}
\caption{\small  Case I-IV: Mean absolute
bias (MAB) and root mean squared error (RMSE) of regression parameter
estimates when $n=250$ } \label{Table4}
\begin{center}
\footnotesize
\setlength{\tabcolsep}{1mm}
\begin{tabular}{c c  cc| ccc | ccc | ccc}
\hline\hline
Sample Size $n=250$ &    &\multicolumn{2}{c}{MLEEM}   &\multicolumn{3}{c}{KDEEM }   &\multicolumn{3}{c}{KDEEM.H } &\multicolumn{3}{c}{KDEEM.LSE } \\
\cline{2-13}
{Error distributions} &    &$MAB$ &$RMSE$      &$MAB$ &$RMSE$ &$RE$      &$MAB$ &$RMSE$ &$RE$      &$MAB$ &$RMSE$ &$RE$\\
\hline

    Case I                          &$\beta_{1,0}$      &15.01&18.65      &16.91&21.34&0.76     &19.03&23.70&0.62     &17.50&21.68&0.74\\
$\bepsilon_1\sim N(0,1)$            &$\beta_{1,1}$      &27.79&34.90      &30.05&37.98&0.84     &36.81&46.14&0.57     &35.03&43.34&0.65\\
\cline{2-13}
$\bepsilon_2\sim 0.5\bepsilon_1$    &$\beta_{1,0}$      &7.45&9.28        &8.39&10.61&0.77      &11.10&13.90&0.45     &12.20&15.55&0.36\\
                                    &$\beta_{1,1}$      &14.85&18.75      &16.45&20.87&0.81     &17.11&21.80&0.74     &17.69&22.45&0.70\\
\hline
       Case II                      &$\beta_{1,0}$      &33.64&42.97      &25.88&35.56&1.46     &42.05&54.74&0.62   &80.13&89.61&0.23\\
$\bepsilon_1\sim U(-3,3)$           &$\beta_{1,1}$      &56.53&70.61      &30.29&40.70&3.01     &65.74&84.36&0.70   &78.68&98.68&0.51\\
\cline{2-13}
$\bepsilon_2\sim 0.5\bepsilon_1$    &$\beta_{1,0}$      &22.70&29.52      &16.54&23.60&1.57     &43.59&48.96&0.36   &69.45&73.51&0.16\\
                                    &$\beta_{1,1}$      &42.06&51.97      &20.49&28.02&3.44     &37.94&50.16&1.07   &80.33&89.44&0.34\\
\hline
        Case III                    &$\beta_{1,0}$                 &27.02&34.24      &15.90&21.04&2.65    &34.94&41.92&0.67   &34.12&44.97&0.58\\
$\bepsilon_1\sim 0.5N(-1.5,0.5^2)+0.5N(1.5,0.5^2)$ &$\beta_{1,1}$  &56.13&69.55      &15.24&19.35&12.93   &42.35&50.08&1.43   &93.80&114.74&0.37\\
\cline{2-13}
$\bepsilon_2\sim 0.5\bepsilon_1$    &$\beta_{1,0}$                 &17.39&21.88      &8.18&10.60&4.27      &13.09&16.47&1.76   &38.82&44.60&0.24\\
                                    &$\beta_{1,1}$                 &36.91&44.91      &7.91&10.01&20.15     &8.92&11.39&15.55    &45.81&55.26&0.66\\
\hline
      Case IV                       &$\beta_{1,0}$                 &44.76&52.12      &39.57&49.34&1.12    &47.99&54.13&0.93   &39.77&45.90&1.29\\
$\bepsilon_1\sim 0.5N(-1,0.5^2)+0.5N(1,1.5^2)$ &$\beta_{1,1}$      &52.02&66.36      &27.47&35.96&3.41    &33.89&46.81&2.01   &39.98&52.20&1.62\\
\cline{2-13}
$\bepsilon_2\sim 0.5\bepsilon_1$    &$\beta_{1,0}$                 &27.92&36.19      &12.72&17.59&4.23    &16.03&20.40&3.15   &29.36&34.25&1.12\\\
                                    &$\beta_{1,1}$                 &50.32&60.89      &16.82&22.19&7.53    &17.01&21.84&7.78   &46.60&55.40&1.21\\
\hline
\hline
\end{tabular}
\end{center}
\end{sidewaystable}

\begin{sidewaystable}[htbp]
\renewcommand{\arraystretch}{1.5}
\caption{\small  Case V-VII: Mean absolute
bias (MAB) and root mean squared error (RMSE) of regression parameter
estimates when $n=250$} \label{Table8}
\begin{center}
\footnotesize
\setlength{\tabcolsep}{1mm}
\begin{tabular}{c c  cc| ccc | ccc | ccc}
\hline\hline
Sample Size $n=250$ &    &\multicolumn{2}{c}{MLEEM}   &\multicolumn{3}{c}{KDEEM }   &\multicolumn{3}{c}{KDEEM.H } &\multicolumn{3}{c}{KDEEM.LSE } \\
\cline{2-13}
{Error distributions} &    &$MAB$ &$RMSE$      &$MAB$ &$RMSE$ &$RE$      &$MAB$ &$RMSE$ &$RE$      &$MAB$ &$RMSE$ &$RE$\\
\hline

    Case V                          &$\beta_{1,0}$      &96.01&108.35    &92.62&100.58&1.16     &77.56&86.56&1.57     &43.75&53.13&4.15\\
$\bepsilon_1\sim \Lambda(0,1^2)$  &$\beta_{1,1}$        &67.90&95.97     &22.36&32.31&8.83      &24.20&43.24&4.93     &57.95&104.06&0.85\\
\cline{2-13}
$\bepsilon_2\sim 0.5\bepsilon_1$    &$\beta_{1,0}$      &39.80&49.88     &21.25&25.71&3.76      &16.26&20.48&5.93     &19.33&25.13&3.94\\
                                    &$\beta_{1,1}$      &52.97&71.19     &10.65&15.75&20.34     &10.20&14.01&25.81    &42.02&59.38&1.44\\
\hline
       Case VI                      &$\beta_{1,0}$      &10.46&13.29    &8.98&11.43&1.35     &12.17&14.93&0.79    &9.88&12.57&1.12\\
$\bepsilon_1\sim Gamma(2,0.5)$      &$\beta_{1,1}$      &20.83&26.44    &12.26&15.74&2.82    &15.77&20.37&1.68    &20.21&25.46&1.08\\
\cline{2-13}
$\bepsilon_2\sim 0.5\bepsilon_1$    &$\beta_{1,0}$      &5.48&6.93      &5.59&7.06&0.96      &11.96&13.93&0.24    &5.53&6.92&1.00\\
                                    &$\beta_{1,1}$      &11.96&15.03    &7.93&10.34&2.11     &9.89&12.86&1.37     &11.24&14.20&1.12\\
\hline
        Case VII                    &$\beta_{1,0}$           &55.29&65.54      &47.52&58.24&1.27    &56.70&67.13&0.95   &74.27&81.90&0.64\\
$\bepsilon_1\sim Rayleigh(3)$ &$\beta_{1,1}$                 &73.80&92.91      &59.89&76.64&1.47    &74.95&93.86&0.98   &83.38&104.17&0.80\\
\cline{2-13}
$\bepsilon_2\sim 0.5\bepsilon_1$    &$\beta_{1,0}$           &24.56&31.46      &22.51&28.79&1.19    &24.40&30.71&1.05   &34.98&40.90&0.59\\
                                    &$\beta_{1,1}$           &51.11&63.07      &44.70&56.14&1.26    &45.69&56.51&1.25   &47.63&59.32&1.13\\
\hline
\hline
\end{tabular}
\end{center}
\end{sidewaystable}

\begin{sidewaystable}[htbp]
\renewcommand{\arraystretch}{1.5}
\caption{\small  Case I-IV: Mean absolute
bias (MAB) and root mean squared error (RMSE) of regression parameter
estimates when $n=500$} \label{Table1}
\begin{center}
\footnotesize
\setlength{\tabcolsep}{1mm}
\begin{tabular}{c c  cc| ccc | ccc | ccc}
\hline\hline
Sample Size $n=500$ &    &\multicolumn{2}{c}{MLEEM}   &\multicolumn{3}{c}{KDEEM }   &\multicolumn{3}{c}{KDEEM.H } &\multicolumn{3}{c}{KDEEM.LSE } \\
\cline{2-13}
{Error distributions} &    &$MAB$ &$RMSE$      &$MAB$ &$RMSE$ &$RE$      &$MAB$ &$RMSE$ &$RE$      &$MAB$ &$RMSE$ &$RE$\\
\hline

    Case I                          &$\beta_{1,0}$      &10.22&12.89      &11.25&14.11&0.83     &12.74&16.08&0.64     &13.17&16.38&0.62\\
$\bepsilon_1\sim N(0,1)$            &$\beta_{1,1}$      &19.14&23.91      &20.17&25.51&0.88     &25.50&32.29&0.55     &24.14&30.60&0.61\\
\cline{2-13}
$\bepsilon_2\sim 0.5\bepsilon_1$    &$\beta_{1,0}$      &5.40&6.80        &5.83&7.34&0.86       &8.24&10.33&0.43      &10.74&13.42&0.26\\
                                    &$\beta_{1,1}$      &11.09&13.75      &11.79&14.82&0.86     &13.17&16.38&0.62     &13.09&16.43&0.70\\
\hline
       Case II                      &$\beta_{1,0}$      &22.09&27.59      &15.28&20.04&1.90     &26.15&34.01&0.66   &74.04&79.69&0.12\\
$\bepsilon_1\sim U(-3,3)$           &$\beta_{1,1}$      &44.78&55.58      &17.75&23.23&5.73     &49.75&63.75&0.76   &58.39&72.61&0.59\\
\cline{2-13}
$\bepsilon_2\sim 0.5\bepsilon_1$    &$\beta_{1,0}$      &18.84&22.80      &10.27&13.29&2.94     &37.80&40.27&0.32   &69.76&71.84&0.10\\
                                    &$\beta_{1,1}$      &34.76&41.81      &11.14&14.54&8.27     &30.79&38.57&1.18   &85.73&90.27&0.21\\
\hline
        Case III                    &$\beta_{1,0}$                 &21.21&26.38      &10.87&13.71&3.70    &29.19&34.62&0.58   &25.15&31.88&0.68\\
$\bepsilon_1\sim 0.5N(-1.5,0.5^2)+0.5N(1.5,0.5^2)$ &$\beta_{1,1}$  &51.42&61.93      &10.33&12.94&22.92   &34.21&47.27&1.72   &84.91&104.96&0.35\\
\cline{2-13}
$\bepsilon_2\sim 0.5\bepsilon_1$    &$\beta_{1,0}$                 &15.35&18.37      &5.95&7.39&6.18      &11.30&13.75&1.79   &36.07&39.71&0.21\\
                                    &$\beta_{1,1}$                 &31.42&37.30      &5.36&6.77&30.33     &6.31&8.01&21.67    &41.85&48.12&0.60\\
\hline
      Case IV                       &$\beta_{1,0}$                 &41.69&46.67      &35.19&43.34&1.16    &47.46&51.19&0.83   &35.50&39.58&1.39\\
$\bepsilon_1\sim 0.5N(-1,0.5^2)+0.5N(1,1.5^2)$ &$\beta_{1,1}$      &41.49&52.11      &18.47&23.05&5.11    &22.78&28.71&3.30   &31.04&38.59&1.82\\
\cline{2-13}
$\bepsilon_2\sim 0.5\bepsilon_1$    &$\beta_{1,0}$                 &25.96&32.73      &9.56&12.73&6.61     &12.25&15.27&4.60   &28..94&31.69&1.07\\\
                                    &$\beta_{1,1}$                 &45.13&53.33      &11.31&14.63&13.28   &12.10&15.30&12.14  &46.60&51.88&1.06\\
\hline
\hline
\end{tabular}
\end{center}
\end{sidewaystable}

\begin{sidewaystable}[htbp]
\renewcommand{\arraystretch}{1.5}
\caption{\small  Case V-VII: Mean absolute
bias (MAB) and root mean squared error (RMSE) of regression parameter
estimates when $n=500$} \label{Table10}
\begin{center}
\footnotesize
\setlength{\tabcolsep}{1mm}
\begin{tabular}{c c  cc| ccc | ccc | ccc}
\hline\hline
Sample Size $n=500$ &    &\multicolumn{2}{c}{MLEEM}   &\multicolumn{3}{c}{KDEEM }   &\multicolumn{3}{c}{KDEEM.H } &\multicolumn{3}{c}{KDEEM.LSE } \\
\cline{2-13}
{Error distributions} &    &$MAB$ &$RMSE$      &$MAB$ &$RMSE$ &$RE$      &$MAB$ &$RMSE$ &$RE$      &$MAB$ &$RMSE$ &$RE$\\
\hline

    Case V                          &$\beta_{1,0}$      &104.26&113.13   &92.64&97.05&1.36     &92.32&98.95&1.31     &39.85&44.29&6.52\\
$\bepsilon_1\sim \Lambda(0,1^2)$  &$\beta_{1,1}$        &62.67&87.65     &20.11&27.61&10.08     &31.48&54.61&2.58    &41.19&59.35&2.18\\
\cline{2-13}
$\bepsilon_2\sim 0.5\bepsilon_1$    &$\beta_{1,0}$      &44.83&51.94     &27.88&31.05&2.80      &24.18&27.19&3.65    &16.50&21.43&5.88\\
                                    &$\beta_{1,1}$      &50.32&66.18     &13.39&18.33&13.04     &11.88&16.97&15.21   &45.40&53.85&1.51\\
\hline
       Case VI                      &$\beta_{1,0}$      &7.66&9.57     &6.34&9.97&1.44     &10.35&12.19&0.62     &7.15&8.91&1.15\\
$\bepsilon_1\sim Gamma(5,1)$        &$\beta_{1,1}$      &15.10&18.87   &7.59&9.53&3.93     &10.27&13.18&2.05     &16.98&20.74&0.83\\
\cline{2-13}
$\bepsilon_2\sim 0.5\bepsilon_1$    &$\beta_{1,0}$      &4.11&5.18     &4.65&5.67&0.83     &12.11&13.33&0.15     &3.98&4.94&1.10\\
                                    &$\beta_{1,1}$      &9.62&12.10    &5.55&7.18&2.84     &7.71&9.83&1.52       &8.00&10.12&1.43\\
\hline
        Case VII                    &$\beta_{1,0}$     &48.95&56.32      &38.61&47.02&1.44    &51.31&58.90&0.91   &73.40&77.61&0.53\\
$\bepsilon_1\sim Rayleigh(3)$       &$\beta_{1,1}$     &63.12&77.70      &43.56&55.11&1.99    &64.28&79.20&0.96   &73.98&90.71&0.73\\
\cline{2-13}
$\bepsilon_2\sim 0.5\bepsilon_1$    &$\beta_{1,0}$     &19.66&25.49      &17.51&22.94&1.24    &19.47&24.92&1.05   &34.12&38.15&0.45\\
                                    &$\beta_{1,1}$     &40.52&50.07      &34.32&43.35&1.33    &36.00&45.62&1.21   &41.65&51.05&0.96\\
\hline
\hline
\end{tabular}
\end{center}
\end{sidewaystable}

\begin{sidewaystable}[htbp]
\renewcommand{\arraystretch}{1.5}
\caption{\small  Case I-IV: Mean absolute
bias (MAB) and root mean squared error (RMSE) of regression parameter
estimates when $n=1000$} \label{Table6}
\begin{center}
\footnotesize
\setlength{\tabcolsep}{1mm}
\begin{tabular}{c c  cc| ccc | ccc | ccc}
\hline\hline
Sample Size $n=1000$ &    &\multicolumn{2}{c}{MLEEM}   &\multicolumn{3}{c}{KDEEM }   &\multicolumn{3}{c}{KDEEM.H } &\multicolumn{3}{c}{KDEEM.LSE } \\
\cline{2-13}
{Error distributions} &    &$MAB$ &$RMSE$      &$MAB$ &$RMSE$ &$RE$      &$MAB$ &$RMSE$ &$RE$      &$MAB$ &$RMSE$ &$RE$\\
\hline

    Case I                          &$\beta_{1,0}$      &7.80&13.74     &8.23&16.14&0.72     &9.94&17.51&0.62     &10.51&16.18&0.72\\
$\bepsilon_1\sim N(0,1)$            &$\beta_{1,1}$      &13.62&19.98    &14.45&25.58&0.61    &18.45&29.40&0.46    &18.15&25.07&0.64\\
\cline{2-13}
$\bepsilon_2\sim 0.5\bepsilon_1$    &$\beta_{1,0}$      &4.10&10.02     &4.13&7.87&1.62      &6.57&10.04&0.99     &11.03&15.19&0.44\\
                                    &$\beta_{1,1}$      &7.93&12.71     &8.02&10.10&1.58     &8.47&10.57&1.45     &9.16&13.94&0.83\\
\hline
       Case II                      &$\beta_{1,0}$      &15.52&19.44    &9.75&12.42&2.45     &16.32&20.68&0.88   &71.51&74.38&0.07\\
$\bepsilon_1\sim U(-3,3)$           &$\beta_{1,1}$      &34.94&42.34    &9.93&12.66&11.19    &35.73&44.23&0.92   &41.75&51.57&0.67\\
\cline{2-13}
$\bepsilon_2\sim 0.5\bepsilon_1$    &$\beta_{1,0}$      &16.55&19.21    &6.79&8.69&4.89      &34.08&35.29&0.30   &70.60&71.68&0.07\\
                                    &$\beta_{1,1}$      &30.70&35.63    &6.48&8.30&18.45     &28.35&33.69&1.12   &90.65&92.80&0.15\\
\hline
        Case III                    &$\beta_{1,0}$                 &15.39&19.21      &8.41&10.46&3.37    &24.57&28.82&0.44   &20.34&25.37&0.57\\
$\bepsilon_1\sim 0.5N(-1.5,0.5^2)+0.5N(1.5,0.5^2)$ &$\beta_{1,1}$  &45.20&51.43      &7.42&9.22&31.11    &24.56&33.70&2.33   &74.26&90.50&0.32\\
\cline{2-13}
$\bepsilon_2\sim 0.5\bepsilon_1$    &$\beta_{1,0}$                 &13.82&15.97      &4.29&5.34&8.95     &10.25&12.22&1.71   &34.33&36.19&0.19\\
                                    &$\beta_{1,1}$                 &29.10&33.11      &3.61&4.55&53.03    &4.56&5.75&33.19    &40.72&44.17&0.56\\
\hline
      Case IV                       &$\beta_{1,0}$                 &41.91&44.53      &35.76&42.36&1.11    &48.64&5073&0.77    &34.32&36.40&1.50\\
$\bepsilon_1\sim 0.5N(-1,0.5^2)+0.5N(1,1.5^2)$ &$\beta_{1,1}$      &33.83&41.81      &12.96&16.58&6.36    &15.93&19.98&4.38   &26.74&32.12&1.69\\
\cline{2-13}
$\bepsilon_2\sim 0.5\bepsilon_1$    &$\beta_{1,0}$                 &27.99&33.30      &8,73&10.90&9.34     &9.67&11.88&7.85    &29.63&31.16&1.14\\\
                                    &$\beta_{1,1}$                 &45.49&52.40      &8.25&10.46&25.08    &9.46&13.73&19.94   &48.44&51.44&1.04\\
\hline
\hline
\end{tabular}
\end{center}
\end{sidewaystable}

\begin{sidewaystable}[htbp]
\renewcommand{\arraystretch}{1.5}
\caption{\small   Case V-VII: Mean absolute
bias (MAB) and root mean squared error (RMSE) of regression parameter
estimates when $n=1000$} \label{Table12}
\begin{center}
\footnotesize
\setlength{\tabcolsep}{1mm}
\begin{tabular}{c c  cc| ccc | ccc | ccc}
\hline\hline
Sample Size $n=1000$ &    &\multicolumn{2}{c}{MLEEM}   &\multicolumn{3}{c}{KDEEM }   &\multicolumn{3}{c}{KDEEM.H } &\multicolumn{3}{c}{KDEEM.LSE } \\
\cline{2-13}
{Error distributions} &    &$MAB$ &$RMSE$      &$MAB$ &$RMSE$ &$RE$      &$MAB$ &$RMSE$ &$RE$      &$MAB$ &$RMSE$ &$RE$\\
\hline

    Case V                          &$\beta_{1,0}$      &109.23&115.24   &95.77&97.98&1.38     &94.41&97.92&1.39     &34.80&38.02&9.19\\
$\bepsilon_1\sim \Lambda(0,1^2)$  &$\beta_{1,1}$        &60.03&81.28     &15.31&21.52&14.27    &23.80&40.64&4.00    &34.22&46.90&3.00\\
\cline{2-13}
$\bepsilon_2\sim 0.5\bepsilon_1$    &$\beta_{1,0}$      &48.93&53.06     &29.25&31.07&2.98     &24.68&26.76&4.01    &13.93&17.97&8.90\\
                                    &$\beta_{1,1}$      &49.10&62.31     &11.13&13.71&20.64    &7.70&10.27&36.78   &42.53&48.00&1.69\\
\hline
       Case VI                      &$\beta_{1,0}$      &6.03&7.40    &4.52&5.71&1.68     &10.17&11.34&0.43     &4.95&6.21&1.42\\
$\bepsilon_1\sim Gamma(5,1)$      &$\beta_{1,1}$        &11.29&14.03   &5.19&6.47&4.70     &7.25&9.11&2.37    &14.32&17.25&0.66\\
\cline{2-13}
$\bepsilon_2\sim 0.5\bepsilon_1$    &$\beta_{1,0}$      &2.82&3.59    &3.88&4.59&0.61     &11.81&12.44&0.08      &2.75&3.45&1.08\\
                                    &$\beta_{1,1}$      &7.34&8.59    &4.13&5.11&3.07     &6.08&7.41&1.46     &5.42&6.85&1.71\\
\hline
        Case VII                    &$\beta_{1,0}$     &46.58&52.06      &31.38&39.36&1.75    &49.39&54.55&0.91   &72.97&75.46&0.48\\
$\bepsilon_1\sim Rayleigh(3)$       &$\beta_{1,1}$     &58.42&69.16      &30.49&40.63&2.90    &56.44&67.83&1.04   &64.53&76.89&0.81\\
\cline{2-13}
$\bepsilon_2\sim 0.5\bepsilon_1$    &$\beta_{1,0}$     &16.32&23.66      &13.76&19.61&1.46    &16.53&22.19&1.14   &35.87&38.84&0.37\\
                                    &$\beta_{1,1}$     &32.35&41.51      &26.36&33.69&1.52    &30.50&38.49&1.16   &41.31&48.87&0.72\\
\hline
\hline
\end{tabular}
\end{center}
\end{sidewaystable}

\section{Real data analysis}
\label{sec:Realdata}

As data collection techniques improve in molecular virology, an
increasing number of data sets were collected and stored, whose
prominent features are mixture and non-normality. These features, if
not approached properly, might result in efficiency loss in
statistical inference. In this section, we evaluate our proposed {\it
  KDEEM } approach by analyzing the
EIAV data set collected from the experiments of Harbin Veterinary
Research Institute (HVRI) conducted by Equine Infectious Disease Research Team in
March 2017.

EIAV is commonly used for Human immunodeficiency virus (HIV) research
because both EIAV and HIV are lentivirus of the retrovirus family,
with similar genomic structure, protein species, infection and
replication style. In March 2017, some Chinese molecular virologists
of HVRI developed a new attenuated
vaccine successfully, which could induce excellent immune protection
and control the spread of EIAV.
Based on the experimental results of Equine Infectious Disease
Research Team, 45 observations were obtained from 8 mixed-gender
horses. All the horses were inoculated with EIAV infectious clone and
5 horses were inoculated with vaccine strain.
The horses were monitored
daily for clinical symptoms, and blood was drawn at regular intervals
(weekly) for assays of platelets, viral replication, sequencing and
virus-specific immune responses.  After the 15 days immunization
period, the five horses inoculated with vaccine (39 observations, ID 1 to 39)
were normal and immune from the virulent strains. The three horses that
were not vaccinated (6 observations, ID 40 to 45 ) had fever and two
of them died at the end of the experiment. To test the immunization
mechanism of the vaccine strains, the outcome variable of interest is the
log value of viral loads, which measures the immune ability of the
infected horses, and the explanatory variables include three
antiviral agents (SLFN11, Viperin and Tetherin), which can be used in
immunodiffusion assay to confirm
wether an animal was protected. Antiviral agents belong to a type of
cell-intrinsic protein which can potentially prevent the virus
intrusion at every step of genes replication. In practice, most
antiviral agents inside a protected animal's body should have
negative effects on the viral loads.

\begin{figure}[htbp]
    \centering
    \includegraphics[width=1.0\textwidth,height=0.4\textheight]{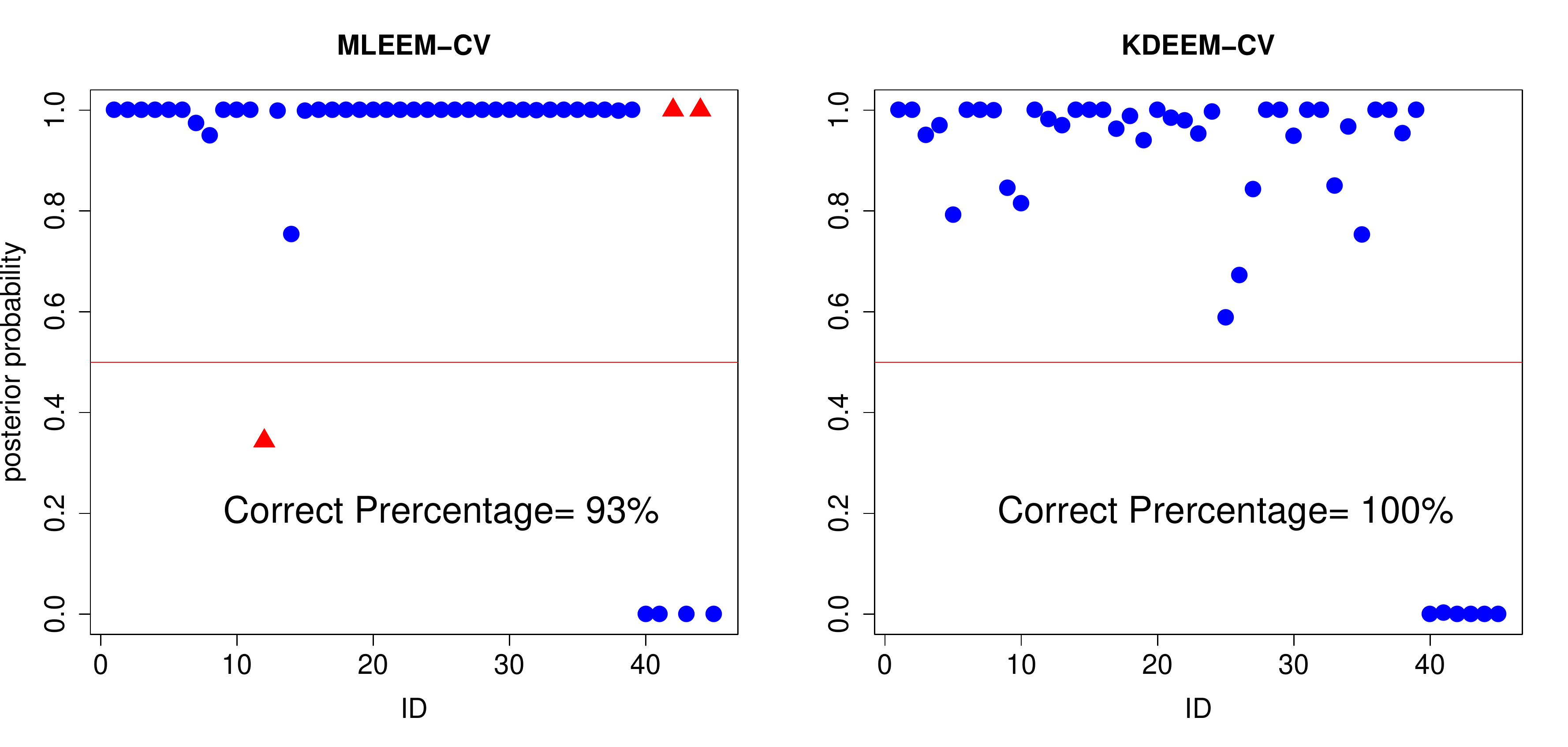}
  \caption{Comparison of the classification error on testing data by the leave-one-out cross validation}\label{figure1}
\end{figure}

We apply the proposed semi-parametric mixture of linear regression
models to help evaluate the lentivirus pathogenesis and immune
protection mechanism.
We obtain the {\it MLEEM } and {\it KDEEM} estimates without using the
vaccine strain information of the horses (or the correlations
among the observations). It is expected that the protected group and
unprotected group might have different relationship between the
response variable and explanatory variables.
Table \ref{Tablerealdata} displays mixture regression parameter
estimates and the correct
  classification percentages (CCP) based on the leave-one-out cross
  validation for the two methods.
  The coefficient estimates indicate that the immunodiffusion mechanisms
  for two groups are significantly different.  For the group of horses
  inoculated with vaccine, both methods  demonstrate that all three
  antiviral agents have negative
  effects on the amount of viral loads inside the animals'
  bodies, which verifies the effectiveness of the vaccine. In contrast, for the
  group of horses that were not vaccinated, two of the three
 antiviral agents have positive effects on the amount of viral loads,
 which is undesirable but sensible since these horses were not
 protected before the experiment.
  The results of CCP
  demonstrate that the new method KDEEM provides more accurate
  classification results than the classical MLE.
  In Figure
  \ref{figure1}, we also plot the classification probabilities that
  the observation is from the protected/vaccinated group versus the ID
  for different estimates. Based on the experiment setup, the
  observations with ID from 1 to 39 belong to the protected group and
  the ones with ID from 40 to 46 are from unprotected group. The red
  triangle points in Figure \ref{figure1} are the observations that
  are wrongly classified. The correct classification percentage
  ($CCP$) of {\it MLEEM } is about 93.33\%, and the CCP of the proposed
  {\it KDEEM} is 100\%. Therefore, the proposed semi-parametric
  mixture of linear regression models can reduce the
  modelling bias and have better classification
  performance for this dataset.

\begin{table}[hbtp]
\caption{{The coefficient estimates and the correct classification
    percentages ($CCP$) based on MLEEM and
    KDEEM. }} \label{Tablerealdata}
\begin{center}
\begin{threeparttable}[b]
\begin{tabular}{c|    c|c|c|c }
\hline
\hline
\multirow{4}{*}{Covariate  }
&\multicolumn{2}{|c|}{\multirow{2}{*}{\tabincell{c}{$MLEEM$}}}
&\multicolumn{2}{|c}{\multirow{2}{*}{\tabincell{c}{$KDEEM$}}}\\
&\multicolumn{2}{|c|}{\nonumber} &\multicolumn{2}{|c}{\nonumber}\\
\cline{2-5}
&\multicolumn{1}{|c|}{\multirow{2}{*}{$\wh{\bb}_{1}$}} &\multicolumn{1}{|c|}{\multirow{2}{*}{$\wh{\bb}_{2}$}}
&\multicolumn{1}{|c|}{\multirow{2}{*}{$\wh{\bb}_{1}$}} &\multicolumn{1}{|c}{\multirow{2}{*}{$\wh{\bb}_{2}$}}\\
&\multicolumn{1}{|c|}{\nonumber}&\multicolumn{1}{|c|}{\nonumber}&\multicolumn{1}{|c|}{\nonumber}&\multicolumn{1}{|c}{\nonumber}\\
\hline
\emph{\multirow{1}{*}{\tabincell{c}{{\bf Intercept}}}}
&6.68 &14.37           &3.50 &11.81       \\
\hline
\emph{\multirow{1}{*}{\tabincell{c}{{\bf SLFN11}}}}
&-0.07 &3.69           &-0.06 &3.75         \\
\cline{2-5}
\emph{\multirow{1}{*}{\tabincell{c}{{\bf viperin}}}}
&-0.33 &-9.13          &-0.42 &-9.25        \\
\cline{2-5}
\emph{\multirow{1}{*}{\tabincell{c}{{\bf Tetherin}}}}
&-0.59  &2.24          &-0.32 &2.26       \\
\hline
\emph{{\bf CCP}}
&\multicolumn{2}{|c|}{93.33\%}&\multicolumn{2}{|c}{100\%}\\
\hline
\hline
\end{tabular}
\end{threeparttable}
\end{center}
\end{table}

%

\section{Discussion}
\label{sec:summary}

Traditional mixture of regression models assume that the component
error densities have  normal distributions, and the subsequent
analysis through MLE will be invalid if the normality assumption is violated.
In this article, we propose a semiparametric mixture regression
estimator with unspecified error densities. We establish the
identifiability of the semi-parametric mixture of
regression models and provide the asymptotic properties of the proposed
estimators. Simulation studies and real data application demonstrate
that the proposed estimators work well for different error densities
and provide substantial improvement over the classic MLE when the
component error densities are non-normal.

To stay focused, we only considered the mixture of linear
regressions. It will be interesting to extend the results in this paper
to some other mixture regression models such as semiparametric mixture
regression models proposed by \citet{huang2012mixture} and
\citet{xiang2018semiparametric} and nonparametric mixture regression
models proposed by \citet{huang13}.
In our semiparametric
regression model (\ref{newmodel}), we assumed that the number of
components is known. It will be also interesting to choose the number
of components data adaptively for (\ref{newmodel}). For parametric
finite mixture models, the information-based criteria methods, such as
AIC and BIC \citep{fraley1998many, keribin2000consistent}, and
hypothesis testing methods \citep{li12,chen12} are commonly used to
choose the number of components. It will be useful to
  adapt the above procedures to the semiparametric mixture model framework.

\clearpage
\section*{References}
\renewcommand{\baselinestretch}{1}
\normalsize
\bibliographystyle{apalike}
\bibliography{YaoMaXuReference,mixture-ref}
\end{document}